\documentclass[conference]{IEEEtran}
\IEEEoverridecommandlockouts

\usepackage{draftwatermark}
\SetWatermarkText{Early Preprint}
\SetWatermarkScale{0.8}

\usepackage{cite}
\usepackage{amsmath,amssymb,amsfonts}
\usepackage{algorithmic}
\usepackage{graphicx}
\usepackage{textcomp}
\usepackage{xcolor}
\usepackage{rotating}
\usepackage{tikz}
\def\BibTeX{{\rm B\kern-.05em{\sc i\kern-.025em b}\kern-.08em
    T\kern-.1667em\lower.7ex\hbox{E}\kern-.125emX}}
\begin{document}

\title{Q2Logic: A Coarse-Grained Architecture targeting Schrödinger Quantum Circuit Simulations}

\author{\textbf{Artur Podobas} \\
\IEEEauthorblockA{\textit{School of Electrical Engineering and Computer Science} \\
\textit{KTH Royal Institute of Technology}\\
Stockholm, Sweden \\
artur@podobas.net}}

\maketitle

\begin{abstract}
Quantum computing is emerging as an important  (but radical) technology that might take us beyond Moore's law for certain applications. Today, in parallel with improving quantum computers, computer scientists are relying heavily on quantum circuit simulators to develop algorithms. Most existing quantum circuit simulators run on general-purpose CPUs or GPUs. However, at the same time, quantum circuits themselves offer multiple opportunities for parallelization, some of which could map better to other architecture-- architectures such as reconfigurable systems. In this early work, we created a quantum circuit simulator system called Q2Logic. Q2Logic is a coarse-grained reconfigurable architecture (CGRA) implemented as an overlay on Field-Programmable Gate Arrays (FPGAs), but specialized towards quantum simulations. We described how Q2Logic has been created and reveal implementation details, limitations, and opportunities. We end the study by empirically comparing the performance of Q2Logic (running on a Intel Agilex FPGA) against the state-of-the-art framework SVSim (running on a modern processor), showing improvements in three large circuits(\#qbit$\geq$27 ), where Q2Logic can be up-to ~7x faster.
\end{abstract}

\begin{IEEEkeywords}
FPGA, Overlay, CGRA, Quantum Simulation, State Vector
\end{IEEEkeywords}

\section{Introduction}

With the end of Dennard's scaling~\cite{bohr200730} and the impending termination of Moore's law~\cite{theis2017end}, scientists are actively searching for alternative models of computation to continue the performance scaling of future computer systems. Today, a plethora of post-Moore alternatives are emerging, such as (for example) neuromorphic computing systems~\cite{schuman2017survey} (that are inspired by the human brain) or reconfigurable systems~\cite{podobas2020survey}. However, out of all emerging post-Moore computing systems, perhaps none is as hyped and disruptive as quantum computing~\cite{gyongyosi2019survey}.

A quantum computer is a computing system that exploits properties of quantum mechanics to perform operations. Similar to a classical computer whose state is best described using bits that can assume the value 0 or 1, a quantum computer similarly has a state described by a number of quantum bits (qubits)~\cite{humble2019quantum}. However -- unlike a classical computer -- a qubit can assume both the values 0 and 1 but also be in a state that is somewhere in-between and which is determined only when measuring said qubit. Needless to say, a quantum computer can operate on a very large computation state, and even a system with 50 qubits (or more) can solve problems that a classical computer simply cannot-- a concept known as quantum supremacy~\cite{arute2019quantum}.

Despite being researched for several decades, quantum computing is still in its cradle, and while there are several \textit{applications} (called quantum circuits) that uses quantum computers, most of these are (relative) small and simple. 
Given the aggressive funding that quantum-related research has been allocated in the years to come, we can expect that field will grow significantly.  And while we can expect larger quantum systems to become available, we will also see a larger need for solid quantum circuit simulation infrastructure due to their versatility to test new ideas, concepts, and/or limitations.

Today, most quantum circuit simulators (e.g., Qiskit~\cite{smelyanskiy2016qhipster} or qHipster~\cite{cross2018ibm}) are executed on general-purpose processors (CPUs) and graphics processing units (GPUs) (e.g., SV-Sim~\cite{li2021sv}). In this work, however, we postulate that by creating Coarse-Grained Reconfigurable Architecture (CGRA)~\cite{podobas2020survey} overlays that can be executed on Field-Programmable Gate Arrays (FPGAs) and that operate on quantum gates and bits (as opposed to classical gates and bits), we can become more performance than running on general-purpose systems. Inspired by earlier work on using FPGAs to accelerate other scientific workloads (e.g.,~\cite{yang2019fully,podobas2017designing,huthmann2019scaling}), this work is our first steps towards faster quantum circuit simulations on reconfigurable architectures. We contribute with the following:

\begin{enumerate}
\item The first (to our knowledge) ever CGRA-like that specializes in accelerating quantum gates and circuits that is both generic and target high qubit circuits, and
\item the design, implementation, and evaluation of our Q2Logic-overlay with common quantum circuits, empirically quantifying the performance gains over state-of-the-methods
\end{enumerate}

The remainder of the paper is structured in the following way: Section II gives a short background on quantum computing, quantum gates, and quantum circuit simulators. Section III introduces Q2Logic, our FPGA overlay for quantum gates, including high-level design, implementation, quantum circuit scheduler, initial performance model, and discussion on constraints and limitations. Section IV shows the results of our design contra state-of-the-art quantum simulators. We end the paper with related work and a conclusion in sections V and VI, respectively.

\section{Background}

The state of a classical computing system is defined by a number of bits and their values. For example, a register in an 8-bit general-purpose processor has a state space that is $2^8$, where the register can be in \textit{only a single of all those 256 states at a certain time}. A quantum computing system is not unlike a classical system, with the exception each state in the system has different probabilities of the system being in that particular state. This is because a quantum bit (qubit) -- unlike a classic bit -- can be in the state of 0, 1, or in state somewhere in-between. The actual state is decided only when an observer measures the state. More formally, the entire quantum system state, $|\psi\rangle$, can be represented as: 

\begin{equation}
|\psi\rangle = \alpha_{0..0}*|0...0\rangle + ... + \alpha_{1..1}*|1..1\rangle   
\end{equation}

where $\alpha$ is a complex value whose amplitude is the probability of the system being in that state. The sum of all probabilities in the system thus equals 100\%.

Modifying the states of a quantum computing system is performed by executing a Quantum Circuit (QC). A QC has a number of horizontal lines, each associated with a particular qubit. Along each horizontal line (each qubit), there can be a quantum gate~\cite{divincenzo1998quantum} that performs an operation on that qubit. Operations can be unary, binary, or even tertiary and thus operate on one or more qubits. Each unary quantum gate, such as the X (Pauli-X or inversion gate), H (Hadamard gate), and I (Identity, do nothing gate), is represented as a 2x2 unary matrix $U$. While there are many binary gates, the most common is CNOT (Controlled-NOT), which is similar to the classic XOR gate in that it takes as input two qubits (one called control) and applies an X-gate on one of the qubits should the control bit be in the '1' state. The CNOT coupled with the unary gates is enough to universally describe any other gate (e.g., Toffoli-gates, etc.). 

There are multiple ways of simulating a quantum circuit using a classical computer, including Feynman paths algorithms~\cite{bernstein1993quantum} or Tensor network contractions~\cite{boixo2017simulation}. However, one of the most widely used algorithms is Schrödinger's algorithm~\cite{de2007massively}, which is the algorithm we chose to work on.

Schrödinger's algorithms require the simulator to store the full state vector of the quantum system, which means that for a qubit system of size $q$, the state vector will occupy $2^{8q}$ bytes (assuming single-precision complex values). Explicitly storing the full state like this means that the amount of space available on the device and its bandwidth is likely to be the limiting factor on the size of the quantum computer being simulated. However, unlike other algorithms, Schrödinger's algorithm place no actual limit on the depth of the circuit, and a Schrödinger-style simulator simply advances the internal state step by step according to the quantum circuit (that is, in polynomial time).

Executing a quantum circuit involves moving along the time axis (from left to right) and applying quantum gates that appear at every single timeslot to their respective qubits. Naively applying each gate $U_{q}$ to qubit $q$ involves computing the Kronecker product of all qubits and matrix multiplying it by the state: 

\begin{equation}
 |\psi\rangle_{t+1} = |\psi\rangle_{t} * I_0 \otimes ... \otimes U_{q} \otimes ... I_n   
\end{equation}

\section{Q2Logic: An Coarse-Grained FPGA Overlay for Quantum Circuits}

Before continuing, we would like to spend a few words on what simulating quantum simulators mean from a computational perspective. \\
\textbf{Implementation}: As should be evident from the parent section, Schrödinger's algorithm focuses on applying small, 2x2 unitary matrixes onto pairs of complex values in the state space. While it is possible to construct a matrix (using the Kronecker product specified in the previous section), implementing the simulation in such a way is impractical since the state grows quadratically to the number of qubits, meaning that the Kronecker matrix would be many times the size of the already problematic size of the state itself. In practice, and as implemented in multiple frameworks~\cite{kelly2018simulating,li2021sv,wu2019full}, we implement the application of a quantum gate (2x2 matrix) on a qubit q by iterating over the entire state, performing a vector-matrix operation using the vector corresponding to the states where the qubit q is 0 and 1 (and all other qubits in identical states in the vector). For example, to apply a gate $U$(2x2 matrix) onto qubit 0 in a 3-qubit quantum computer, we would multiply and update the following states: 

\begin{center}
\begin{equation}
\begin{bmatrix} |000\rangle \\ |001\rangle \end{bmatrix},
\begin{bmatrix}|010\rangle \\ |011\rangle \end{bmatrix},
\begin{bmatrix}|100\rangle \\ |101\rangle \end{bmatrix},
\begin{bmatrix}|110\rangle \\ |111\rangle \end{bmatrix}
\end{equation}
\end{center}

Similarly, if we instead wanted to apply a gate U onto qubit 1 in a 3-qubit quantum computer, we would multiply and update the following states instead: 

\begin{center}
\begin{equation}
\begin{bmatrix}|000\rangle \\ |010\rangle \end{bmatrix},
\begin{bmatrix}|001\rangle \\ |011\rangle \end{bmatrix},
\begin{bmatrix}|100\rangle \\ |110\rangle \end{bmatrix},
\begin{bmatrix}|101\rangle \\ |111\rangle \end{bmatrix}
\end{equation}
\end{center}

The observant reader will notice a similarity between said above access pattern and the typical butterfly networks when computing Fast Fourier Transforms (FFTs)~\cite{cochran1967fast}. \\

\textbf{Locality}: Understanding the above access patterns, we can also understand that the algorithm has ample opportunities to exploit both temporal and spatial data locality. Assume, for example, that in a quantum circuit time step, we perform two quantum gate operations on two different qubits. For the sake of simplicity, let's assume we perform operations on qubit \#0 and \#1. We know that we will be reading the entire state space twice, giving us the opportunity to store intermediate results from our first state-space read to immediately compute the operation on qubit \#1. Furthermore, in the case of qubit \#0, there is near-perfect spatial locality, as we are always the next state. \\

\textbf{Parallelism}: The algorithm offers multiple forms of parallelism. One opportunity is to divide the entire state space into chunks of size ($2^n$), where each processor (or system) operates inside one chunk. Another opportunity is pipelined parallelism in order to exploit the temporal locality properties of the algorith. Here, one system or processor  consumes $|\psi\rangle_{t}$ (where $t$ is the state before a particular qubit operation) and produces $|\psi\rangle_{t+1}$, where a second system or processor consumes $|\psi\rangle_{t}$ and produces $|\psi\rangle_{t+1}$. This type of parallelism is not easily exploited on traditional von-Neumann systems, but are more favorable to reconfigurable systems.

\subsection{The Q2Logic Architecture}

\begin{figure*}
\includegraphics[width=1\textwidth]{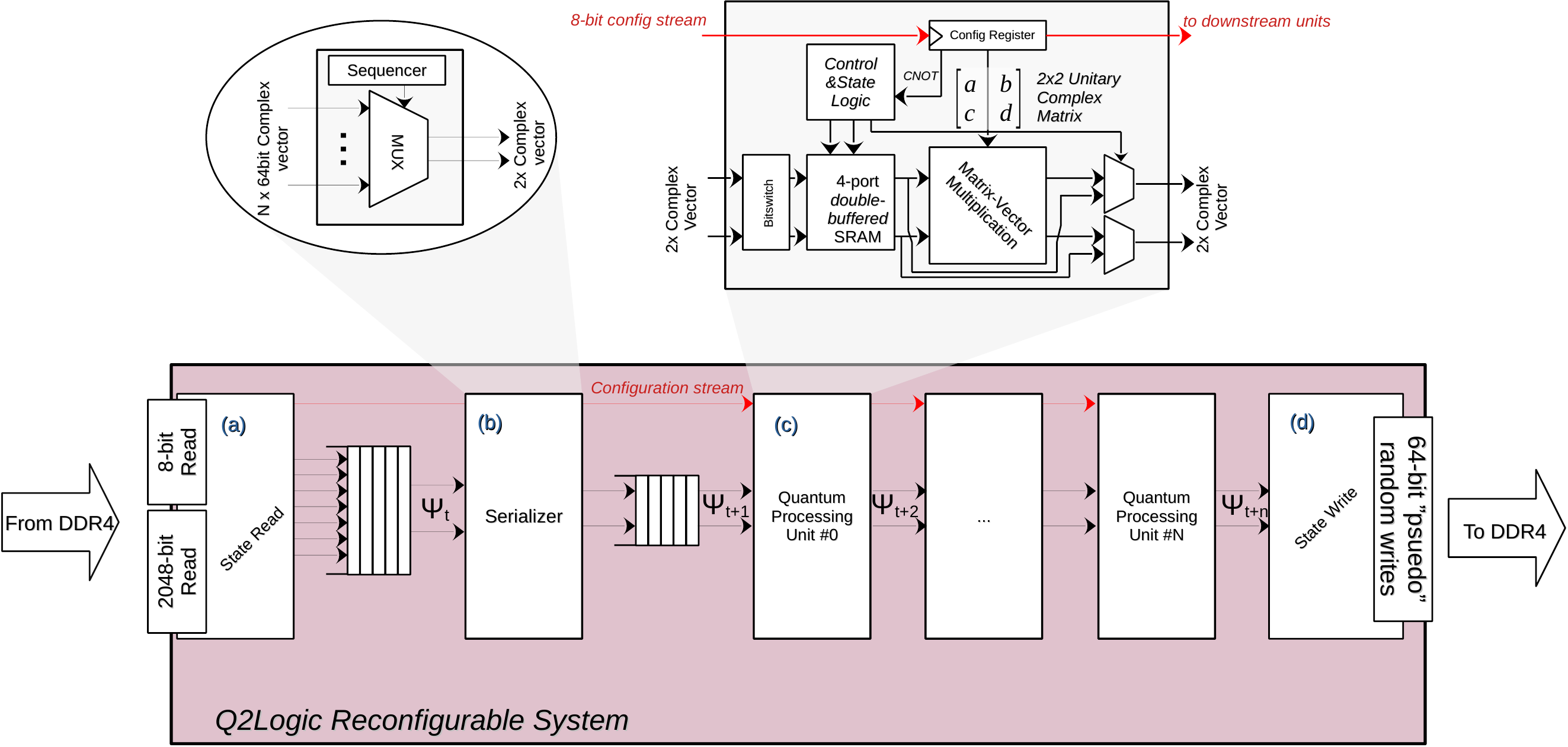}
\caption{The Q2Logic architecture and its various components, including: (a) external address-generators for efficiently fetching the quantum computer state $|\psi\rangle$, (b) a serializer for transforming wide burst-reads into into a smaller serial stream, (c) quantum processing units capable of performing a single quantum gate (unary or CNOT) on the quantum state (including internal components), and (d) external address-generator for writing back the $|\psi\rangle$ and projecting it to a different mapping.}
\label{fig:architecture}
\end{figure*}

We introduce the Q2Logic system, which is a coarse-grained architecture capable of exploiting the multiple forms of parallelism and locality indicated above. The overall Q2Logic system is shown in Figure~\ref{fig:architecture}, and we will in detail describe each of the internal components of the system. The system was described in the OpenCL programming model, using the Intel OpenCL SDK for FPGAs. 

\subsubsection{State-Reader (a)}: The state-reader component is responsible for two parts: \textbf{(i)} configuring the system with a bitstream and \textbf{(ii)} fetching the full quantum computer state $|\psi\rangle$. When the system is booted, the state reader starts by reading the bitstream. Each QPU in the system is represented by a 44-byte configuration, where the configuration decide the type of QC the system should perform. The state-reader will read the entire bitstream and stream it serially through all the QPUs along an 8-bit interface. Upon completion, the state-reader will fetch the entire state $|\psi\rangle$ and stream it through the system. The state-reader has very wide, burst-coalesced access to external memory since the quantum state is read linearly without any strides.

\subsubsection{Serializer (b)}: the serializer is implemented to decouple the very wide, burst-coalesced accesses that the state-reader performs (and which are needed for decent external read memory bandwidth performance) and the narrow internal vectors that are streamed through the QPUs. 

\subsubsection{Quantum Processing Unit (QPU, c)}: the QPU is the heart of the system and is the component that applies a particular gate to a particular qubit. The QPUs divide the incoming state space into chunks (of varying sizes, subject to design-time decisions) that are operated on inside the QPUs. The operation of the QPU is described as follows:

\begin{enumerate}
\item A stream of complex values (quantum states) is loaded into the QPU. Each value written is labeled with a number in order that it arrived: the first value has a label of 0, the second a label of 1, the third a label of 2, and so forth. 
\item The QPU takes the above label and switches the bit according to the qubit that we want to compute on and bit 0 in the label. The QPU then stores the value in the SRAM at the local of the new (bit-switched) label. 
\item Once the SRAM is full with values, the computation begins, and the QPU reads the SRAM content sequentially, fetching two complex values at a time and performing a vector-matrix operation, outputting the results,
\item If the quantum gate is to be applied a CNOT, then the control-bit will drive a multiplexer to either output the newly computed value, or output the previous states' value.
\end{enumerate}

One important note is that it is the size of the internal SRAM that dictates how many different qubits (in series) we can process. For example, if the internal SRAM has a capacity of $2^{4}$ complex values, then we can (at most) compute on four qubits in series in the quantum circuit (e.g., qubits \#0-3 or qubits \#6-9), while if we have a capacity of $2^{16}$, we can compute of up-to 16 different qubits in the quantum circuit (e.g., qubits \#5-\#20). We call the number of qubits that a particular configuration can handle as $N_{sysqbits}$. As we will see later, this property is important for scheduling QCs onto the Q2Logic system.

The above operation holds true with one exception: instead of dividing the operation into a load and compute phase, we instead double the internal SRAM and perform the load and compute phase in parallel. Finally, note that we can place multiple QPUs in serial, where each QPU stream its output to the input of the next.

\subsubsection{State-Writer (d)}: The state-writer consumes the final stream of complex values from the last QPUs and writes them back to memory. Similar to the QPUs, we label each incoming value in the order of arrival. This label is the address to which we will write the value in external memory. Before writing it back, we rotate (left or right) said address. This rotation is performed to prepare how values are placed in the memory depending on which qubits the next bitstream will use.

For example: if we have a quantum circuit with 16 qubits and have configured our QPUs to be able to work with at most four qubits in sequence. Let's assume we first want to work with qubits \#0-3, and in the next bitstream, we want to work with qubits \#8-11. In such a scenario, when executing the first bitstream, we would make sure to transform the write-back address by rotating the bits 5 steps to the left and then launch the next bitstream. 

Important to note that, unlike the state reader, the state-writer does not always write back data to memory in a nice (well-behaved) coalesced way. The performance is depending on the rotation, which have different memory performance (see Section~\ref{sec:perf_consid} for more information).

\subsubsection{Configuration Bitstream}: similar to any other reconfigurable device, such as a CGRA or FPGA, the Q2Logic system is configured with a bitstream that decides system operation. The configuration bitstream and interface are used prior to using the device, where the configuration bitstream contains information about the quantum circuit to be executed, the unitary matrixes to be applied to the quantum states, the type of quantum gates to be applied (unitary or CNOT) and how to transform address-mappings between processing units.
The configuration stream, internally, is an 8-bit wide bus that runs through the accelerator, starting from the state reader and ending with the last processing unit in the chain. Implementation-wise, since the accelerator is described in OpenCL, we use a shift register to move the configuration through the accelerator; upon finishing the configuration, we access said shift register through a union in order to access the different sub-components of the configuration (the unitary matrix, address mappings, etc.)

\begin{figure*}[h]
\begin{center}
\includegraphics[width=0.8\textwidth]{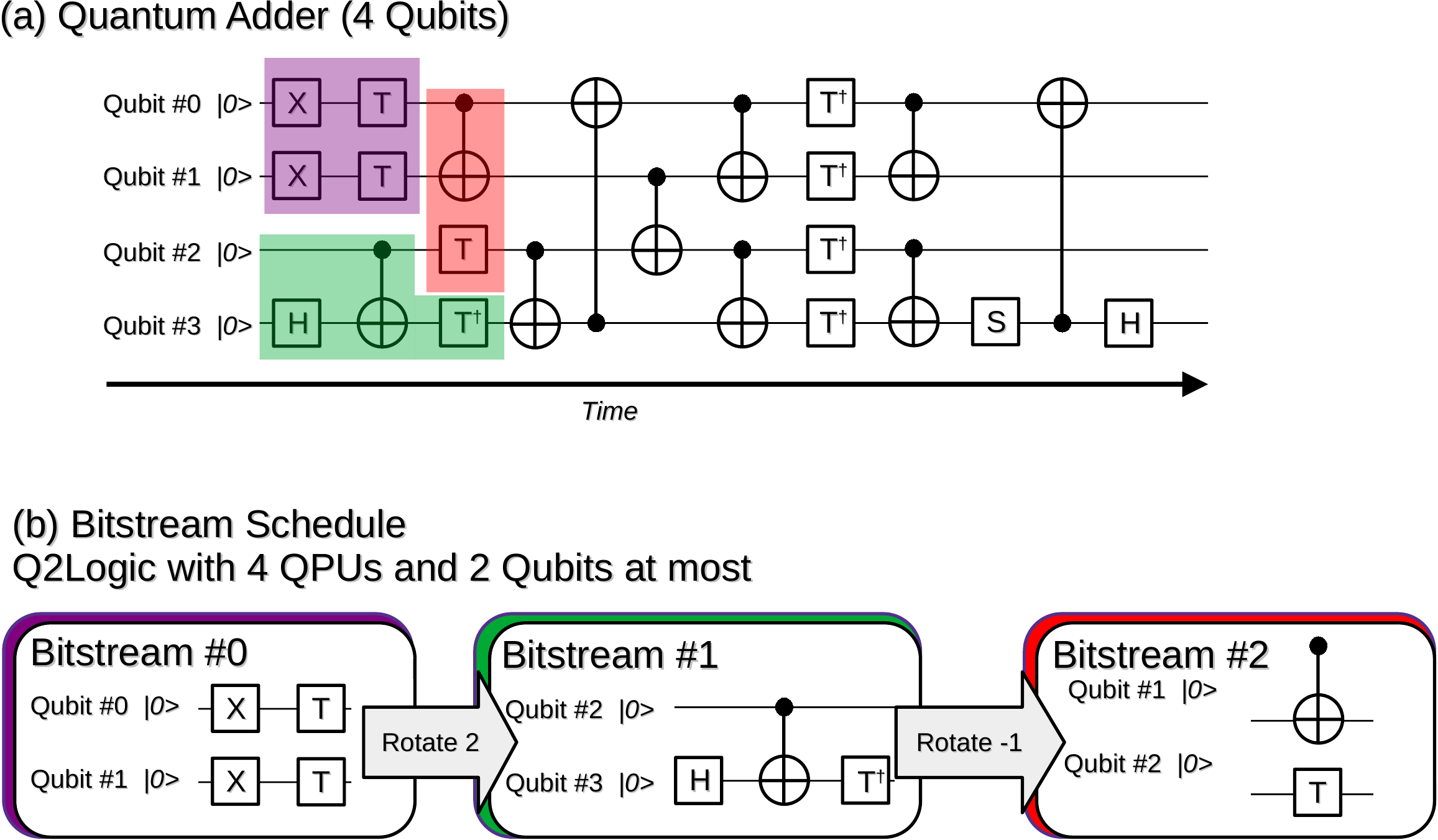}
\end{center}
\caption{An example quantum circuit (a quantum adder) shown in \textbf{(a)}, and a three bitstreams (out of many) showing how the quantum circuit was scheduled on-to a small, hypothetical Q2Logic system \textbf{(b)}.}
\label{fig:circuit}
\end{figure*}

\subsubsection{Mapping Quantum Circuits on Q2Logic}
To illustrate how our Q2Logic system work, we will briefly show how we schedule a part of a full quantum circuit. We will be focusing our attention on the quantum circuit of an adder, which consists of 4 qubits, shown in Figure~\ref{fig:circuit}:a. Now, let us assume that we have a Q2Logic system that is capable of executing at most four qubit gates (that is, it has 4 x QPUs) and that it is capable of at most computing on two consecutive qubits (that is, the SRAM inside each QPU holds four complex values). Scheduling our adder circuit onto such a Q2Logic system for the three first bitstreams is shown in Figure~\ref{fig:circuit}:b. We start with the first bitstream, which we decide will operate on qubits \#0 and \#1, and manage to schedule 2 x Pauli-X and 2 x T gates, fully utilizing our Q2Logic system. For the next bitstream, we notice that all existing gates have a dependency on the H-gate and CNOT gate operating on qubits \#2 and \#3. Hence,  the second bitstream is chosen to operate on qubits \#2-3, and the first bitstream is modified to rotate the writeback address by two steps, placing the states of qubits \#2-3 where the qubits \#0-1 previous where in memory. The third bitstream follows a similar logic, scheduling the T-gate and a CNOT-gate operating on qubits \#1-2, implying that we inserted a -1 rotation into the second bitstream. This effort continues by creating more bitstreams while honoring system constraints and gate dependencies.

\subsubsection{Performance Considerations}
\label{sec:perf_consid}
Understanding the system and the example of scheduling a circuit onto the Q2Logic system, we now turn our attention to the performance considerations of our system. There are two items that govern performance: \textbf{(i)} the Q2Logic system performance and \textbf{(ii)} the effectiveness of the circuit scheduler.

\textbf{Q2Logic Performance Considerations:} The time to run a bitstream (part of a quantum circuit) on Q2Logic is governed by five variables: \textbf{(i)} the number of QPUs, $N_{QPU}$, \textbf{(ii)} the width of the burst-coalesced loads, $B_{lw}$ (bits) in state-read (up-to 2048-bit wide), and \textbf{(iii)} the operating frequency, $f_{max}$ (in MHz) in comparison to the peak memory frequency clock, $f_{mem}$, \textbf{(iv)} the configuration overhead, $t_{cfg}$, and \textbf{(v)} a tunable parameters $\alpha$ that is a function of the rotation. The execution time to execute a single bitstream, $t_{bitstream}$ is on a state space $8*|\psi\rangle$ (bytes, assuming each state is a complex value in IEEE-754 single-precision) approximates:

\begin{equation}
t_{bitstream} = t_{cfg} + \frac{8*size(|\psi\rangle)}{\alpha * min(f_{mem},f_{max}) * (1 + \frac{B_{lw}}{128} } 
\end{equation}

\begin{figure}[h]
\begin{center}
\includegraphics[width=0.9\columnwidth]{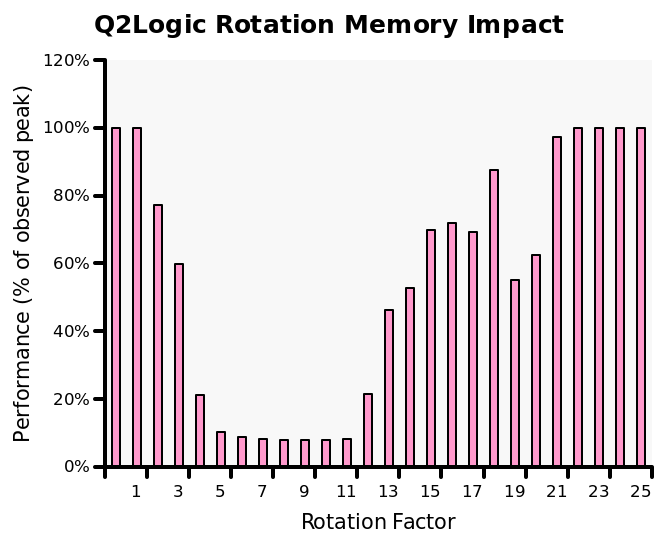}
\end{center}
\caption{The impact of rotation the address-bits of a stream when writing back, showing that some rotations are more favorable than others.}
\label{fig:rotationperformance}
\end{figure}

The $\alpha$ variable is important since it has to empirically be found through extensive testing of stream bandwidth (see, e.g.,\cite{zohouri2019memory}) to find the fraction of peak that the system can achieve under ideal conditions, multiplied by the impact of rotation of the address bits during writeback (which is a function of the quantum circuit scheduling). To give an example of the impact of the rotation factor on the bandwidth of the Q2Logic system, see Figure~\ref{fig:rotationperformance}. Here, we performed controlled experiments on a sample Q2Logic system with 16 QPUs, isolating the impact of different rotations while recording the bandwidth of the system. We notice that depending on the rotation factor, a performance penalty of nearly ~12x can be observed. This penalty kicks in when rotation between 4 and 12 bits), likely coming from how the external memory banks (four in our cases) are interleaved, leading to serialization of writebacks.

\textbf{Circuit Scheduler}: The second performance consideration is the circuit scheduler, which is a NP-Complete problem. The goal of the scheduler is to minimize the number of bitstream configurations that we need to cycle through in order to finish our quantum circuit. For the circuit scheduling to be effective, we want to maximize the number of qubits, $N_{sysqbit}$ that our Q2Logic can internally handle-- the higher $N_{sysqbit}$, the more freedom the scheduler has to find parallel quantum gates to schedule. However, there is a limit on the $N_{sysqbit}$, which is often the limit on the number of on-chip resources (SRAM) that the underlying architecture has. Finally, even if the scheduler does a good job at utilizing Q2Logic QPUs, if the needed rotations between bitstreams happen in a poor-performing region, then it may preferably use a less-dense but still better-performing schedule.

\textbf{Our circuit scheduler} attempts is a greedy-based scheduler that tries to maximize the efficiency of the schedule while taking the rotation penalty into mind. While it works well, there is still room for improvement, and we will consider an integer linear programming (ILP) or constraints-based programming (CSP) solution in the future.

\section{Results and Experimental Platform}

The Q2Logic system was created using Intel OpenCL SDK for FPGAs version 22.2 and synthesized with Quartus 22.2. The target FPGA platform was the DE10-Agilex platform, featuring an Intel Agilex FPGA AGFB014R24B2E2V1 and four DDR4 banks (32 GB in total) of memory. All designs were compiled with the following options: \texttt{-global-ring -high-effort -ffp-reassoc -ffp-contract=fast -ffp-reassociate}.

\begin{table}
    \centering
    \begin{tabular}{c | c | c | l  }
     \# Circuit & \# Qubits & \# Gates/CNOTs & Description \\ \hline \hline
          Adder          &     28   &  424 / 195   &  Quantum adder\\           
          Cat\_state     &     28   &  28 /27 &  Cat-state ($|0...0>$ or $|1...1>$)\\
          Ising          &     28   &  302 / 54   &  Ising model simulation\\
          Multiplier     &     20   &  1079 / 464  & Multiplication \\
          Knn            &     25   &  230  / 96    &  Quantum K-nearest neighbors \\
          QFT            &     27   &  1782 / 702 &  Quantum fourier transform\\
          Swap\_test     &     25   &  230 / 96      &  Simple test\\
          WState         &     27   & 157 / 52 & WState preparation \\
          VQC            &     27   & 4560 / 1456 & Variational Quantum Classifier \\        
    \end{tabular}
    \caption{Benchmarks used in the study, including their \# qubit count, the number of gates, and description.}
    \label{tab:benchmark_overview}
\end{table}

The quantum circuit benchmarks that we used are summarized in Table~\ref{tab:benchmark_overview} and were generated by the NWQ benchmark~\cite{li2021qasmbench}. We intentionally generated fairly large circuits (most use 25+ qubits) since we see little meaning in simulating anything less (that will generally run within milliseconds on a general-purpose system). The QASM benchmarks were converted to our internal format, and bitstreams were generated. 

We compared the performance of our Q2Logic system against the SV-Sim~\cite{li2021sv} simulator running on a state-of-the-art workstation with a 12th generation i9-12900K, 128 GB RAM, and running Ubuntu 12.04 LTS. SV-Sim was configured with OpenMP enabled and with single-precision (\texttt{using ValType = float} in config.hpp) support. 

Altough the performance comparison favour the i9-12900k compared to the DE10-Agilex in terms of characteristics, they do share many similarities. Both have identical external memory bandwidth (76.8 GB/s).

For all benchmarks, we used all eight of the performance cores of the i9-12900K when running with SV-Sim, and the best-performing Q2Logic system for that particular application. We exclude intialization and bitstream generating activities (which, in some cases, can be significant-- in particular bitstream generating for larger quantum circuits).

\subsection{Synthesis results}
The synthesis results for our Q2Logic system on the DE10-Net Agilex platform is shown in Table~\ref{tab:synthesis}. Here, we have varied how many QPUs the system has, $N_{QPU}$, and how many adjescent qbits, $N_{sysqbits}$, we can compute on within the bitstream. Overall, we note that our system operates using a fairly high frequency, between 172-228 MHz, even on those design that occupy a large amount of FPGA resources (e.g., $N_{QPU}=48$ or $N_{QPU}=64$). With that being said, ideally, we would reach a clock frequency of 300 MHz, since that would allow us to better use the external memory bandwidth, so there is some room for future critical path optimizations. 
We also note that the OpenCL high-level synthesis (HLS) tool instantiates the same number of internal BRAM resources independent of whether we set $N_{systqbits}$ to 6, 8, or even 10, and only when $N_{systqbits}=12$ does the system consume more BRAM resources. Practically, this means that for our target FPGA, it is never worth going for a $N_{sysqbits}$ lower than 10 since it does not bring any benefits. There are two \textit{extremes} of our designs: (i) the system that has $N_{QPU}=48$ and focuses on $N_{sysqbits}=14$, which is BRAM-bound, and (ii) $N_{QPU}=48$ and focuses on $N_{sysqbits}=12$, which focuses on larger number of gates, and is logic-bound. The first, (i), makes it easier to schedule (since we can consume a larger number of bits) while the second encourage fewer bitstreams. Using either of these, is -- as we shortly will see -- very application-dependent. Finally, there is still ample opportunities for better use DSPs, since our largest desing uses ~37\% of the DSPs resources (amounting to a peak performance of ~857 GFLOP/s).

\begin{table}
    \centering
    \begin{tabular}{c  | c | c | c | c | c }
     \# $N_{QPU}$ & $N_{sysqbits}$ & ALMs & RAM Blocks & DSPs   & $f_{max}$ \\ \hline \hline
          16      &     6          &  \textbf{34\%} &    13\%  &  6\%    &   224.46 MHz  \\ 
          16      &     8          &  35\% &    13\%  &  6\%    &   227.89 MHz \\
          16      &     10         &  35\% &    13\%  &  6\%    &   \textbf{228.46} MHz \\
          32      &     6          &  48\% &    15\%  &  11\%   &   215.7 MHz  \\
          32      &     8          &  49\% &    15\%  &  11\%   &   194.81 MHz \\
          32      &     10         &  49\% &    15\%  &  11\%   &   199.08 MHz \\
          32      &     12         &  50\% &    24\%  &  11\%   &   219.58 MHz \\
          32      &     14         &  52\% &    62\%  &  11\%   &   200.64 MHz \\
          48      &     14         &  67\% &    87\%  &  17\%   &   202.14 MHz \\
          64      &     10         &  \textbf{78\%} &    19\%  &  23\%   &   \textbf{172.47 MHz} \\
          64      &     12         &  \textbf{80\%} &    \textbf{37\%}  &  23\%   &   180.44 MHz \\          
    \end{tabular}
    \caption{Synthesis results for different Q2Logic systems}
    \label{tab:synthesis}
\end{table}

\subsection{Performance}

\begin{figure*}[h]
\includegraphics[width=1.0\textwidth]{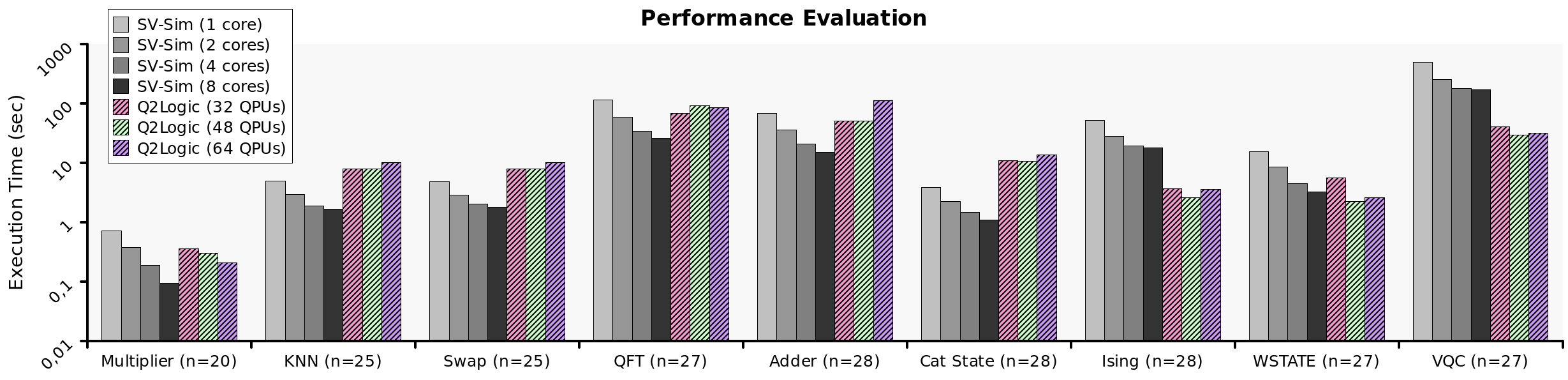}
\caption{Performance comparison between Q2Logic (Agilex) and SV-Sim (i9-12900K) showing time to solution (y-axis) of various large quantum circuits (x-axis).}
\label{fig:results}
\end{figure*}

The absolute performance of the Q2Logic system, when compared to the SV-Sim software, is shown in Figure~\ref{fig:results}. Here, we see the performance of the different quantum circuits (x-axis) running on the different systems with the execution time (y-axis) showing. A lower execution time is better.

Overall, the FPGA system is competitive with the SV-Sim simulations. For smaller quantum circuits or sparse circuits (Multiplier, KNN, Swap), the SV-Sim performs better, especially when all eight cores are used. In many cases, the Q2Logic system is on par (or better) than running SV-Sim with 1 or 2 cores but loses out when more core counts are used. For some circuits, e.g., the QFT circuit, our scheduler is unable to find a good schedule that fully utilizes are QPUs, leading to sub-par performance. The largest difference that is in favor of SV-Sim is the Cat-state (a toy benchmark) circuit, which is nearly 10x the difference.

However, for some circuits where the circuit scheduling works well, such as Ising, Wstate, and VQC, we see a rather significant improvement in using the Q2Logic over the SV-Sim system. For instance, for VQC (which is a dense circuit), we see an improvement of 5.75x between the best Q2Logic and best SV-Sim simulation in favor of Q2Logic. Similarly, we see an improvement of 6.92x and 1.44x for Ising and Wstate, respectively, both in favor of Q2Logic. 

\begin{figure}[h]
\includegraphics[width=.9\columnwidth]{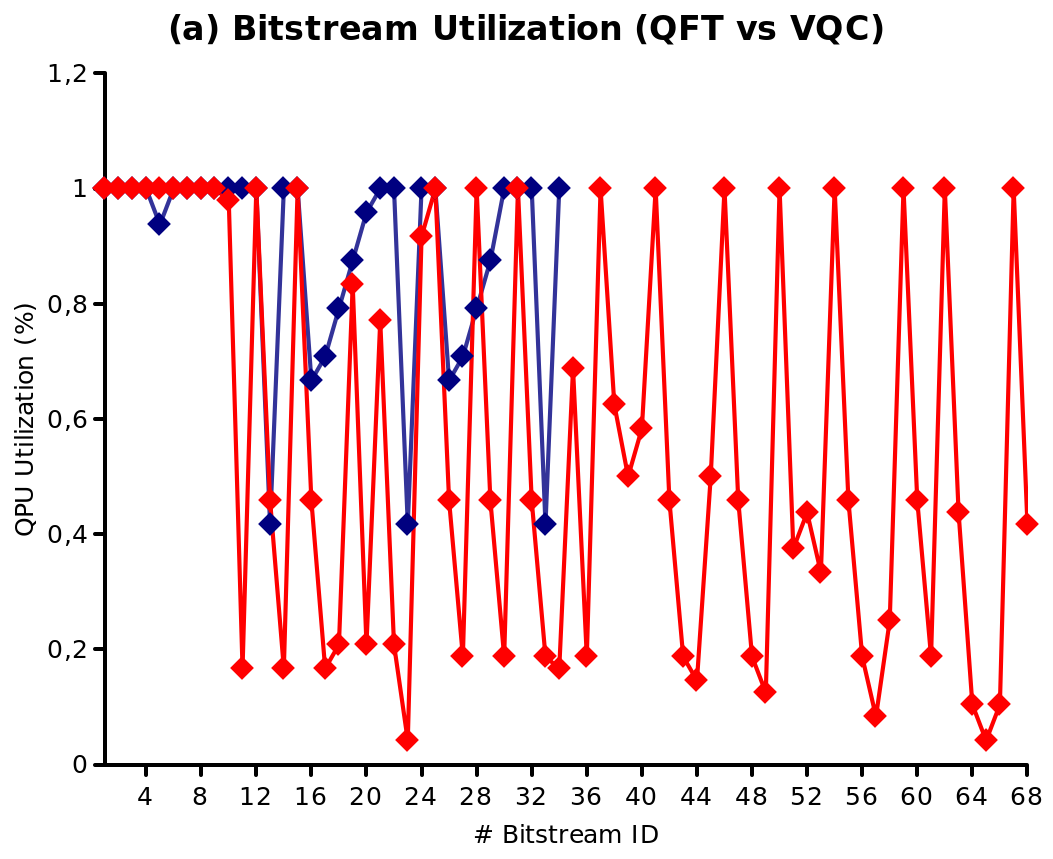}
\includegraphics[width=.9\columnwidth]{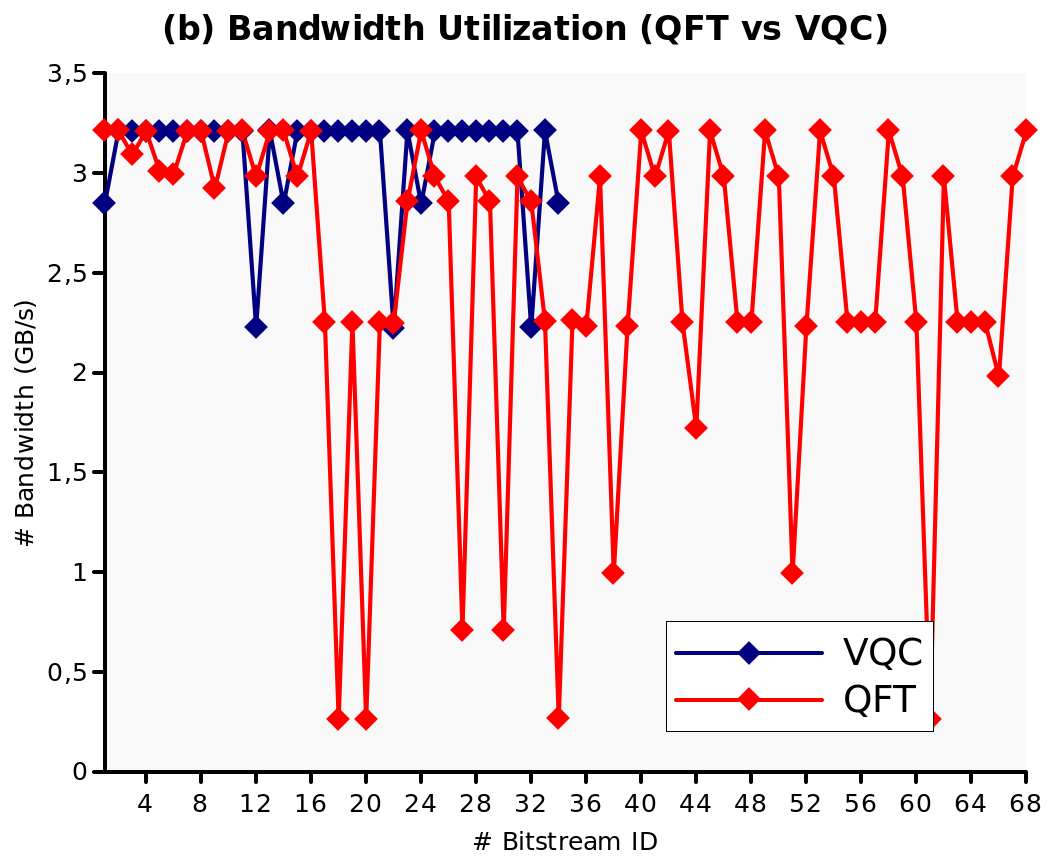}
\caption{Example showing a good schedule and a poor schedule (number of bitstreams to execute the circuit is on the x-axis) between two different circuits, where (a) shows how the VQC vs QFT circuit on resource utilization, and (b) shows the memory bandwidth.}
\label{fig:sample}
\end{figure}

To further illustrate how important a good circuit scheduler is, we compare QFT and VQC-- two circuits with same number of qubits (27) . We seen in Figure~\ref{fig:sample}:a that the VQC oinly needs 34 bitstreams to execute, and most of the bitstreams are full (near 100\% with some caveats), while the QFT circuit requires 68 bitstreams due to poor utilization. Furthermore, our scheduler manage to optimize to relatively good memory bandwidth for the VQC application in Figure~\ref{fig:sample}:b, operating at near peak for the system, while it fails to do the same for the QFT circuit, leading to sub-par performance.

\section{Related Work}

To the best of our knowledge, our Q2Logic system is the first to provide a generic (not application- or circuit-specific) acceleration of relatively large (20+ qubits) quantum circuits.

Early efforts, such at those by Khalid et al.~\cite{khalid2004fpga},  mapped QFT and  Grover's search (3 qubits) onto FPGAs. The accelerators executed at ~80 MHz and showed that their accelerators is nearly three orders of magnitude (1000x) better than the corresponding CPU implementation (libquantum~\cite{glendinning2004parallelization}). One limitation of the study is that the authors used very small quantum circuits (3  qubits), which took microseconds to execute. 

Recently, Hong et al.~\cite{hong2022implementation} mapped both Groover's algorithm (3 qubits) onto an FPGA and QFT~\cite{hong2022quantum}, showing performance improvements over CPU (Groover's 3-qbit is 62x faster on the FPGA) and their QFT is faster than prior FPGA work.

The work by Pilch~\cite{pilch2019fpga} showed the Deutsch Algorithm algorithm working on an FPGA and also explored the impact that increased qubits have on the hardware, emulating up to 5 qubits.

Some work has specialized on a particular creating a specialized accelerator for a particular quantum circuit. In particular, the QFT seems popular and has been the subject of some studies. Qian et al. ~\cite{qian2019efficient} accelerated QFT on FPGAs and showed between 1-2 orders of magnitude in improvements over software for a 4-qbuit and 8-qbiut system, respectively. The work by Waidyasooriyta~\cite{waidyasooriya2022scalable} is another such work, that is very exciting as it -- unlike many other works -- focuses on high qubit simulations. They showed 28-30 qubit QFTs reaching between ~20x  and ~80x  faster compared to state-of-the-art frameworks such as QuEST~\cite{jones2019quest} and HpQC~\cite{bian2020hpqc}.

There are other algorithms for simulating quantum circuits have also been mapped on FPGAs, e.g.,. Tensor networks~\cite{levental2021tensor}, but these have different characteristics and limitations compared to the algorithm used in this paper and will thus not be discussed further.

\section{Conclusion}
We have presented (to our best knowledge) the first reconfigurable system for a generic, high qubit (most are 25+) quantum circuit. Our system, called Q2Logic, belongs to the CGRA-like systems where the unit of reconfiguration are matrixes (2x23). Furthermore, we discussed implementation details and decisions, developed an initial model describing performance, and described key challenges with the architecture. Finally, we compared our system on an Intel Agilex FPGA against a state-of-the-art CPU and framework (SV-Sim), demonstrating that our framework -- albeit preliminary -- is capable of reaching up to 7x performance on dense, high-qubit quantum circuits. 
In the future, seeing the large impact the circuit scheduler has on performance, we intend to pursue a more formal and better scheduler. Furthermore, there are ample opportunities to improve the performance of the architecture, including parallel execution of streams (to improve bandwidth) and optimizations such as BRAM double-pumping~\cite{canis2013multi} (which, for some reason, does not work on our platform). We expect that applying the above optimization, along with achieving a 300 MHz clock frequency, will yield another ~3-5x performance improvement over what we empirically measured today.

\bibliographystyle{IEEEtran}
\bibliography{ref}

\end{document}